\documentclass[lnicst,12pt]{svmultln}

\makeatletter
\renewcommand\subsubsection{\@startsection{subsubsection}{3}{\z@}%
                       {-18\p@ \@plus -4\p@ \@minus -4\p@}%
                       {4\p@ \@plus 2\p@ \@minus 2\p@}%
                       {\normalfont\normalsize\bfseries\boldmath
                        \rightskip=\z@ \@plus 8em\pretolerance=10000 }}
\renewcommand\paragraph{\@startsection{paragraph}{4}{\z@}%
                       {-12\p@ \@plus -4\p@ \@minus -4\p@}%
                       {2\p@ \@plus 1\p@ \@minus 1\p@}%
                       {\normalfont\normalsize\itshape
                        \rightskip=\z@ \@plus 8em\pretolerance=10000 }}
\makeatother

\usepackage{etex}
\usepackage{colortbl}
\usepackage{amsmath}
\usepackage{txfonts}
\usepackage[ansinew]{inputenc}
\usepackage[T1]{fontenc}
\usepackage{lmodern}  \normalfont
\usepackage{amsfonts}
\usepackage{ae,aecompl}
\usepackage{booktabs}
\usepackage{array}
\usepackage{tikz}
\usepackage{tabularx}
\usepackage{listings}
\usepackage{longtable}

\usepackage{geometry}
\usepackage{makeidx}  

\usepackage{graphics}
\usepackage{graphicx}
\usepackage{url}
\usepackage{lscape}

\usepackage[leftcaption]{sidecap}
\usepackage[labelsep=endash, textfont={footnotesize, singlespacing}, margin=10pt, format=plain, labelfont=bf]{caption}
\newcommand{\fakeparagraph}[1]{\smallskip\noindent\textbf{#1.}}

\usepackage{multirow} 
\usepackage{tabularx}
\usepackage{listings}
\usepackage{longtable}
\usepackage{booktabs}
\usepackage{color}
\usepackage{hyperref}
\usepackage{tikz}
\usepackage{algorithmic}
\usepackage{algorithm}

\definecolor{name}{rgb}{0.5,0.5,0.5}
\definecolor{javared}{rgb}{0.6,0,0} 
\definecolor{javagreen}{rgb}{0.25,0.5,0.35} 
\definecolor{javapurple}{rgb}{0.5,0,0.35} 
\definecolor{javadocblue}{rgb}{0.25,0.35,0.75} 

\rmfamily
\DeclareFontShape{T1}{lmr}{bx}{sc}{<->ssub * cmr/bx/sc}{}
\DeclareSymbolFont{calletters}{OMS}{cmsy}{b}{n}
\DeclareSymbolFontAlphabet{\mathcal}{calletters}
\DeclareSymbolFont{rmletters}{OMS}{ptm}{m}{n}
\DeclareSymbolFontAlphabet{\mathrm}{rmletters}

\usepackage{epsfig, floatflt, amssymb} 
\usepackage{multirow} 
\usepackage{url} 
\usepackage{textcomp} 
\usepackage[right]{eurosym}
\usepackage{setspace} 
\usepackage{subfig}

\usepackage{xkeyval}
\usepackage{amsgen}
\usepackage{etoolbox}
\usepackage{supertabular}
\usepackage{datatool}

\begin{document}

\mainmatter              

\title{A Testbed for  Experimenting  Internet of Things Applications}

\author{Parthkumar Patel\inst{1} \and Jayraj Dave\inst{1} \and 
Shreedhar Dalal\inst{1} \and Pankesh Patel\inst{2} \and Sanjay Chaudhary\inst{1}}
%

\tocauthor{ }

\institute{
School of Engineering and Applied Science,Ahmedabad University\\
\email{parth.p.btechi13@ahduni.edu.in},
\email{jayraj.d.btechi13@ahduni.edu.in},
\email{shreedhar.d.btechi13@ahduni.edu.in},
\email{sanjay.chaudhary@ahduni.edu.in}\\
\and
ABB Corporate Research,India\\
\email{pankesh.patel@in.abb.com}
}

\maketitle            
\begin{abstract}       
The idea of IoT world has grown to multiple dimensions enclosing different technologies and standards which can provide solutions and goal oriented intelligence to the widespread things via network or internet. In spite of different advancement in technology, challenges related to assessment of IoT solutions under real scenarios and empirical deployments still hinder their evolvement and significant expansion. To design a system that can adequately bolster substantial range of applications and be compliant with superfluity of divergent requirements and also integrating heterogeneous technologies is a difficult task. Thus, simulations and testing to design robust applications becomes paramount elements of a development process. For this, there rises a need of a tool or a methodology to test and manage the applications. This paper presents a novel approach by proposing a testbed for experimenting Internet of Things (IoT) applications. An idea of an open source test bed helps in developing an exploited and sustainable smart system. In order to validate the idea of such testbed we have also implemented two use cases.
\end{abstract}

\section{Introduction}

\textquotedblleft 19th century was a century of empire, the 20th century was a century of nation states and the 21st century will be a century of cities.\textquotedblright \\
\hspace*{4.7in}- Former Denver Mayor W. Webb
\\\\
\hspace*{7mm} Cities are the eventual fate of mankind, with over half of the world\textquotesingle s population now living in urban areas ~\cite[Harrison and Donnelly, 2011]{fifteen}. By 2050, the UN predicts this number will increment to 70\% because of huge economic development in the current urban regions  ~\cite[Nations, 2010]{sixteen}. Some of this development will be in 27 mega cities with population of 10 million individuals, though more than this half of the development will be in urban areas that have less than 500,000 individuals. ~\cite[Naphade et al.,2011]{seventeen}. This urban development and relocation are putting more emphasis on city foundation and bringing up challenges in domains including transportation, energy, security and health care. ~\cite[Haubensak, 2011]{eighteen}.

\begin{figure}[h]
\centering
\includegraphics[width=0.7\linewidth,height=0.25\textheight]{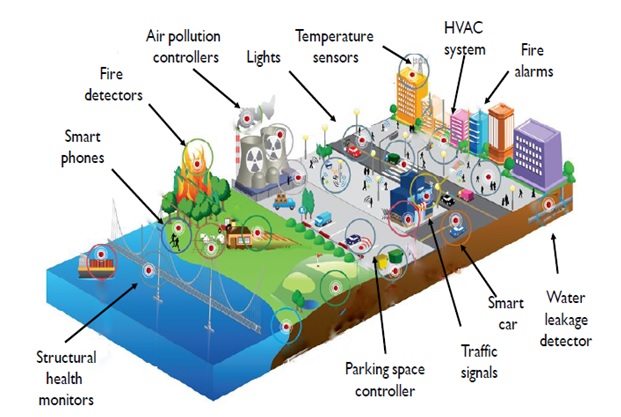}
\caption{Smart objects in urbanized areas(figure credit : http://www.libelium.com)} 
\label{fig:figure1}
\end{figure}
\hspace*{7mm} A viable solution of the aforementioned challenges may be in modern technological advancement in ICT (Information and Communication Technology). These advancements have been engendering an enormous development in the number of smart devices (or things) ~\cite[Vasseur and Dunkels, 2010,p. 3]{nineteen}.  In 2010, the number of everyday physical objects was around 12.5 billion. Cisco predicates that this number is expected to twofold to 25 billion in 2015 as the number of smart devices per individual increases, and to a further 50 billion by 2020 ~\cite[Evans, 2011]{twenty}. Figure ~\ref{fig:figure1} describes the electronic skin of the city comprising of a huge number of smart objects: smoke detectors, temperature sensors, lights, smart phones, fire alarms, air pollution controllers, parking space controllers, and so on. These smart devices sense the physical world by getting data from sensors, influence the physical world by prompting actions utilizing actuators, connect with clients by collaborating with them at whatever required, and process captured data and impart it to outside world.
\\\\
\hspace*{7mm}As a way to implement the above vision, the Internet of Things (IoT) enables varied physical items or things , for example, sensors, actuator, cell phones, and so forth to communicate among them and coordinate with their neighbours to achieve a mutual objective  ~\cite[Atzori et al., 2010]{twentyone}. In spite of the fact that the exact definition of the IoT is as yet advancing, we construct our work in light of the definition beneath, proposed by the CASAGRAS extend: \footnote{http://www.grifs-project.eu/data/File/CASAGRAS\%20FinalReport\%20(2).pdf }:
\\\\
\hspace*{7mm}\emph{\textquotedblleft The Internet of Things is a global network infrastructure, linking physical and virtual objects through the exploitation of data capture and communication capabilities. This infrastructure includes existing and evolving Internet and network developments. It will offer specific object-identification, sensor and connection capability as the basis for the development of independent cooperative services and applications. These will be characterized by a high degree of autonomous data capture, event transfer, network connectivity and interoperability. \textquotedblright }
\\\\
\hspace*{7mm} In the IoT, things procure knowledge on account of the way that they access data that has been accumulated by other devices. For instance, a building cooperates with its inhabitants and neighbouring buildings in situation of fire for safety and security of occupants, workplaces alter themselves automatically as per clients inclinations while reducing energy utilization, or traffic signals control in-stream of vehicles as indicated by the current traffic status
~\cite[de Saint-Exupery,2009]{twentytwo}.
\\\\
\hspace*{7mm} As apparent above, IoT applications will include associations among divergent gadgets, a large number of them connect with their physical environment. An important challenge that should be tended is to minimize the efforts of the partners involved in developing application without impacting the pace of IoT development. 
The above-mentioned challenges have been researched in the related fields of Wireless Sensor Networks (WSNs) ~\cite[Vasseur and Dunkels, 2010,p.11]{nineteen} and ubiquitous computing ~\cite[Vasseur and Dunkels, 2010,p.7]{nineteen}, and perceived as predecessors to IoT. While the fundamental challenge in the previous is the large-scale usage of devices whereas latter suffers from the problem of heterogeneity of devices and the interaction of user with it. (cf. the exemplary smart home situation where a client controls lights and gets notifications from his fridge and toaster.) The goal of the project is to develop such interconnected IoT applications. 



\subsection{TECHNICAL CHALLENGES}

\textbf{Lack of division of roles}: IoT application development is a multi-disciplined process where learnings and concerns from various domains intersects. Conventional IoT application development take the assumption of all the individuals engaged in the application development process have similar potential and set of skills. This is in clear clash with the shifted set of abilities required in the procedure, deployment specific knowledge , domain expertise , platform specific understanding, and application design and deployment knowledge, a challenge identified by recent works, such as~\cite[Chen et al., 2012;Picco, 2010]{twentythree}. 
\\\\
\textbf{Heterogeneity}: IoT applications execute on a system comprising of heterogeneous devices regarding their types, interaction modes and requirements, as well as various platforms and frameworks. The heterogeneity generally spreads into the application code and makes the convey ability of code to an alternate deployment scenario troublesome. Preferably, a similar application should perform equally efficiently and effectively on an alternate implementation environment (e.g., the same home automation application should execute on different setup with various smart devices).
\\\\
\textbf{Scale}: As mentioned above, IoT applications execute on distributed frameworks or platforms comprising of hundreds to thousands of devices, including the efficient and seamless interaction of their activities. Requiring the capability of thinking at such levels of scale is unreasonable by and large, as has been to a great extent the view in the WSN community. Thus, there is a need of sufficient reflections that enable stakeholders to express their requirements in a compact manner without the concern of the scale.
\\\\
\textbf{Different life cycle stages}: Stakeholders need to address issues that are ascribed to various life cycles stages including design, prototyping, development, testing and deployment. At the development stage, the application logic must be observed and decoupled into a set of distributed tasks for underlying network comprising of an expansive number of heterogeneous devices. At the deployment phase, the application logic has to be deployed, tested and verified onto a large number of devices. At the organization stage, the application logic must be deployed onto particular platform of a device. Manual efforts in above phases for thousands of heterogeneous devices is tedious and error prone process. Ideally, there should be system which allows automation with significant reduction in manual efforts by stakeholders.

\subsection{Contribution}
\hspace*{7mm} Due to the challenges, discussed above, the IoT application development is very costly in terms of time and efforts for people with a limited technical expertise (such as domain experts or people with a limited programming expertises). In addition to that the wide reach and higher heterogeneity of the IoT devices suggests crucial challenges in application development catering the emerging real world problems. Here, Testbed provides an open, dynamic, stable, and secure environment enabling easier application testing and design. We have designed a testbed that can aid people to validate their ideas quickly before they deploy in the real world environment, thus improving innovations.The objective of these testbed is to design and implement experimental environments that will grant:
\\\\
\hspace*{12mm} 1.The technical analysis and assessment of IoT solutions under real scenarios\\
\hspace*{12mm} 2.The evaluation of social adoption of novel IoT solutions, and\\
\hspace*{12mm} 3.The magnitude of service usability, potential, efficiency and effectiveness with users in the development process.\\\\
\hspace*{7mm} Test bed is an environment which is configured in accordance to meet the identified test goal for the application/product. Test beds are needed to understand a framework for experimentation of new technologies under realistic operating conditions. It is a development environment (or an environment set-up) which is useful for conducting rigorous and replicable testing of solutions, concepts, theories and new technologies.Test beds are developed to realize the actual scenarios with the freedom of wide experimental possibilities (Contrary to actual scenarios which are more confined in nature).  A test bed is a platform which allows:\\
\\
\hspace*{12mm} 1. Experimentation (Use of different technologies)\\   
\hspace*{12mm} 2. Evaluation (various use cases)\\
\hspace*{12mm} 3. Testing of service usability (interactive visualization)\\ 	
\hspace*{12mm} 4. Testing of technical stability
\\\\
\hspace*{7mm} In the context of IoT and Big Data Analytics,As mentioned the technical challenges above, Any business (or research) looking for IoT implementation have some concerns regarding cost, Efficiency, performance, etc. The idea of testbed helps in resolving these concerns prior to the actual implementation.IoT based Big Data system requires storage and processing infrastructure (Expensive), Sensors deployment, maintenance, etc.  Test bed helps in reducing the time and cost of making this infrastructure and make the development much faster and easier. Due to the factors such as interoperability and difference of latency and storage among different technologies, it becomes very imperative to use the loop feedback control system. Once the system is deployed, it may become complicated to modify  the implemented solutions. Therefore, a development of testbed is of paramount importance for developing fairly large and complex system with integrating multiple technologies from IoT, big data, cloud computing, etc.Using a Testbed one can:\\\\
\hspace*{12mm} 1. Implement specific user scenarios (Traffic, Energy Saving, Healthcare, etc.)\\
\hspace*{12mm} 2. Experiment with connecting different technologies together in a non-production environment\\
\hspace*{12mm} 3. Build final project requirements using test results\\
\hspace*{12mm} 4. Discover opportunities for products and services outside the initial scope
\\\\
\hspace*{7mm} The idea of this generic test bed is meant to be an open source project so that it allows   users to freely modify the implementation of testbed with using various technologies. This open source project will remove barriers between innovators and promote free exchange of ideas within a community to drive scientific and innovative technological advancement. As developer community keeps contributing to such projects it becomes less prone to bugs and more secure than the proprietary systems. Especially, in the field of IoT and big data analytics, many alternative technologies are used for specific contextual challenges and opportunities. Thus, the choice of open source is the key to enable interoperability with other technologies, businesses, research works, etc. As this project is open to innovators and  research community, there will be many possibilities of developing new derivative solutions to real world problems using the similar test bed. The idea of having this project as open source will  significantly reduce the cost and time involved in development of such end-to-end systems. For example, in this paper we have developed two use cases using the same flow and architecture. As claimed above, the development of second use case was much easier to develop.
\\\\
\hspace*{7mm} We have evaluated our testbed on two scenarios:  The first use case is about traffic scenario, where using the multiple emerging technologies visualization has been generated which converts raw data into useful knowledge. In the second use case, real motion detection sensors are deployed to enable the smart lighting in the building. In this use case, the system is analysing the energy consumption of the lighting set up in the building. Due to better opportunities, the world is observing an increasing trend of rural population migrating to cities. In order to meet the massive inrush of rural population into urban location, the cities need to emerge as smart cities. The National Smart Cities mission of India has allotted a budget of \$160 billion for the coming five years to develop 100 smart cities across the country. 
\\\\
\hspace*{7mm} The increasing need for providing better lifestyle in smart cities, has forced the civil authorities to look upon economic prosperity, environment sustainability  and social well being. The increasing urban population raises issues on traffic congestion and delays. A congested roads leads to issues such as fuel-wastage, longer waiting time, air and noise pollution,etc.  In order to have smart tranport management in cities, one of the prerequisites is to have smart ICT infrastructure that is dynamic, open, platform independent and handle real time big data. In order to develop such large scale system, development  of testbed becomes useful for consistent and rapid progress.  Thus, the first use case for a testbed is of traffic scenario where simulated real time traffic data is accumulated and convert this raw data into useful knowledge. This knowledge enables the smart decision making for the stakeholders and can be further developed for smart recommendations and suggestions.
\\\\
\hspace*{7mm} Also, the increasing mass in urban areas demands for better and optimal  management of resources (Especially, energy). In smart cities, big infrastructure requires humongous amount of energy. It raises the issue of underutilization of 24x7 running devices and components which consumers lot of energy. Thus, it creates a need to optimize the use of these resources and control the energy consumption in an efficient way.  

\subsection{Outline}
\hspace*{7mm} The remainder of this chapter is organized as follows: Section ~\ref{sec:proposedtestbed} presents the proposed testbed. Section ~\ref{sec:implementation} presents an implementation of our proposed testbed. Section ~\ref{sec:evalution} evaluates the proposed testbed. Section ~\ref{sec:relatedwork} reviews related work. Finally, Section ~\ref{sec:conclusion} discusses conclusion and future directions.

\section{Proposed Testbed}\label{sec:proposedtestbed}
\hspace*{7mm}  From the analysis of requirements of a testbed, it appears that architecture can be centralized with closely coupled heterogeneous set of sensing nodes deployed over different required locations generate divergent that that are to be delivered to storage and processing layer with the use of suitable communication protocols. A primary characteristics of generic test bed is its ability of incorporating various technologies with available communication infrastructure to collect, store and process data and further to be analyzed to provide novel functions and services. Another prime aspect of this is to make the extracted knowledge easily accessible to an end user for further awareness and realization of the system. 
\subsection{Generic Architecture}
\begin{figure}[h]
\centering
\includegraphics[width=\linewidth]{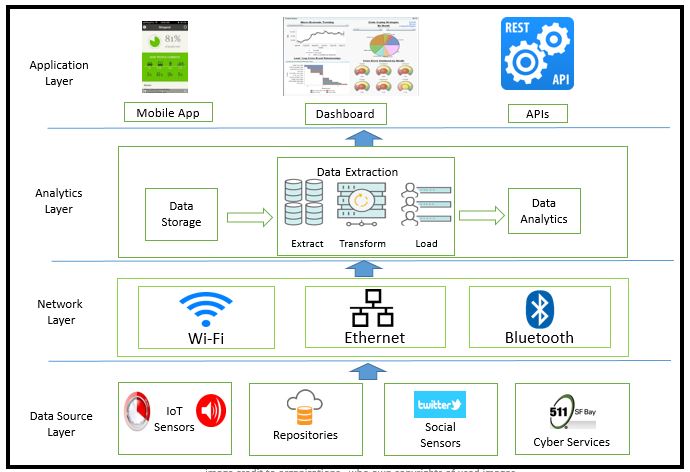}
\caption{Generic Architecture of IoT based Big data analytics end-to-end system}
\label{fig:figure2}
\end{figure}

\hspace*{7mm} In this paper, we are proposing a generic testbed which can successfully inculcate various domains. Figure ~\ref{fig:figure2} illustrates the proposed testbed which is a generic end-to-end system consisting of different layers such as data source layer (perception layer), network layer (Machine-to-Machine communication protocols), analytics layer (data storage and processing) and application layer (visualization).
\\
\fakeparagraph{\emph{Data Source Layer}}: It comprises of the physical items, sensors, social sensors like Twitter and  facebook, Dataset shared by government authorities, cyber services like 511.org. This layer basically deals with the gathering of objects specific data from above mentioned sources. Contingent upon the kind of sources, the data can be about temperature, motion, location, humidity, tweets from social media and etc. The gathered data from the data sources is then passed to Network layer so that it can securely transmit the data to the analytics layer.
\\\\ 
\fakeparagraph{\emph{Network Layer}}: The part of networking layer is to associate all data sources together and allow data sources to share the information with other connected systems. Additionally, the networking layer is capable for conglomerating data from existing IT foundations~\cite[Internet of things in industries: A survey]{fifth}. It securely transfers the collected data from data sources to the data analytics layer. The transmission medium for transferring data can be wireless or wired and technology can be Wifi, Bluetooth, Ethernet etc. depending upon the data source. 
\\\\
\fakeparagraph{\emph{Analytics Layer}}: The Internet of Things will produce a massive data of smart objects. Consequently, it is important to consider how to oversee information of IoT successfully and how to execute online analytical questions and processing conveniently. This layer is capable for the administration and it can connect to the database. Analytics layer gates the data from Network layer which was transmitted by data sources and it stores the received data in the database. The data have to be stored and used shrewdly to take further decisions. It performs data processing techniques and ubiquitous computation and takes programmed choice based on the results. This layer helps in augmenting row data to valuable information sought to satisfy need of an IoT application.
\\\\
\fakeparagraph{\emph{Application Layer}}: This layer is an important layer for IoT application as it acts as an interface to interact with application for end users. Technologies like smartphone, computers, tablets are used as a medium to make IoT application interactive. This layer plays an important role to convert extracted knowledge into visualization. For a layman to completely profit by the IoT transformation, alluring and easy to understand visualization has to be made. This will likewise empower strategy creators to convert information into knowledge which is crucial in quick decision making. This includes both identification of events and visualization of the related crude  information, with information represented to as per the requirements of the end user.

\subsection{Technology used to implement Architecture}
1. \textbf{MQTT} : MQTT is an ISO standard lightweight, machine-to-machine (M2M), publish-subscribe based protocol which is used on top of the TCP/IP protocol ~\cite[MQ telemetry transport]{sixth}. It is intended to deploy technology at remote locations with limited network bandwidth. MQTT protocol needs a message broker, which is responsible for distributing messages to subscriber based on a topic of a message.
\\\\2. \textbf{Kafka} : Kafka is publish-subscribe based stream processing system which has in built partitioning, fault tolerance, better throughput, replication. This makes it more preferable messaging system for real time streaming data ~\cite[Kafka, Apache]{seven}. It uses  Zookeeper to monitor the coordinations among broker, producers and consumer. Here, Kafka is one of the best alternative for real time big data streaming where  multiple consumer subscribe to a topic under which message passing occurs.
\\\\ 3. \textbf{R} : It is an open source programming language which provides software environment for computational statistics and graphics ~\cite[R language definition]{eight}. We have used shiny package from RStudio to build interactive visualization with R.
\\\\ 4. \textbf{Node.js} : Node.js is an open source, cross platform scripting language used at server side which uses javascript runtime environment for executing code on server. It is designed such that commands execute in parallel, and sends callback when signal completion or failure occurs.
\\\\5. \textbf{MySQL} : MySQL is an open source database management system supporting relational schema with structured query language interface. Here, we have used it to stores the data collected from IoT sensors.

\section{An implementation of our proposed testbed}\label{sec:implementation}
\subsection{Use Case - 1: Traffic }
\hspace*{7mm} In order to realize feasibility of the proposed testbed, traffic data simulator is implemented with the help of our testbed, which is one of the use cases for the proposed testbed.  There is legitimate issue of authenticate traffic data for the simulation and research purposes. Thus, analyzing and observing many available traffic data sets ~\cite[Virginia Department of Transportation]{twelve} ~\cite[CityPulse Dataset Collection]{thirteen} ~\cite[OPEN DATA AARHUS]{fourteen}, we have generated simulated traffic readings for the development of a traffic scenario use case. 
The dataset description is mentioned in Table ~\ref{table1}:
\begin{figure}[h!]
\centering
\includegraphics[width=1\linewidth,height=0.4\textheight]{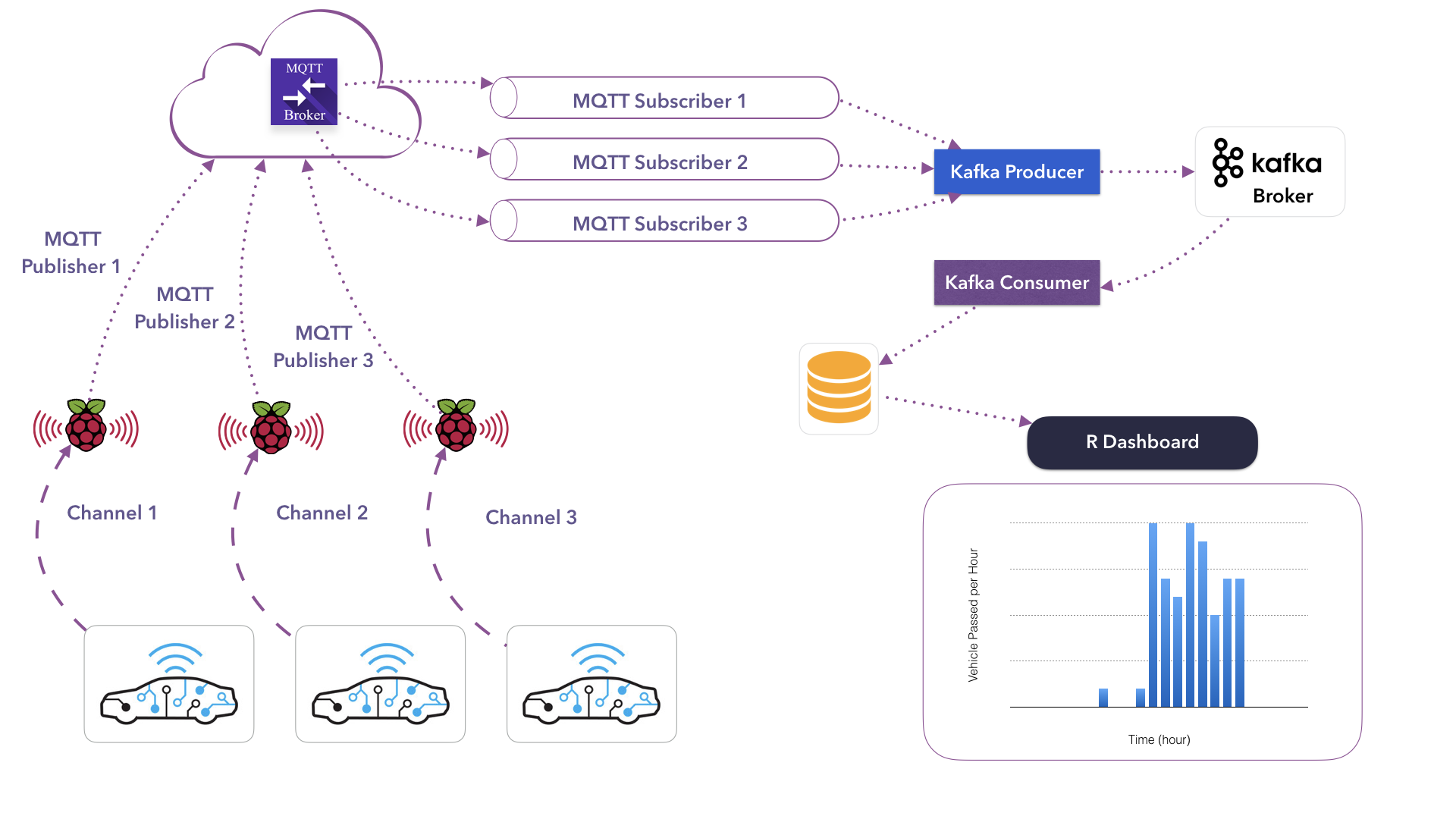}
\caption{Implemented Architecture of Traffic Scenario}
\label{fig:smart_traffic}
\end{figure}
\begin{table}[]
\centering
\begin{tabular}{|c|c|}
\hline
\textbf{Attribute} & \textbf{Description} \\ \hline
Sensor ID & Sensor id linked with a link id \\ \hline
Timestamp & Time of observation taken \\ \hline
2WMV & No. of two-wheeled motor vehicles observed \\ \hline
CarV & No. of Cars and taxis observed \\ \hline
BusV & No. of Buses and Coaches observed \\ \hline
LGV & No. of light goods vehicle observed \\ \hline
HGV & No. of heavy goods vehicle observed \\ \hline
HGVR2 & No. of two-rigid axle HGVs observed \\ \hline
HGVR3 & No. of three-rigid axle HGVs observed \\ \hline
HGVR4 & No. of four or more rigid axle HGVs observed \\ \hline
HGVA3 & No. of three or four articulated axle HGVs observed \\ \hline
HGVA5 & No. of five-articulated axle HGVs observed \\ \hline
\end{tabular}
\caption{Dataset Description of Traffic Scenario}
\label{table1}
\end{table}

\hspace*{7mm} Simulated data (assumed as real time traffic data) is being generated through node.js script. Simulated traffic data is being published by MQTT publisher located on top of the raspberry pi module. This published data is sent to the remotely located MQTT broker. Now, An MQTT subscriber on another remotely located server subscribes to the data located in the queue of MQTT broker. These real time simulated data is sent through Kafka producer to Kafka Consumer. Kafka is feasible option for real time streaming of  big data.  Kafka consumer inject this data into the database which further is used for visualization purposes. R platform uses this database to request queries for visualization on dashboard. A package named Shiny is used along with R for interactive dashboard.
\subsection{Use Case - 2: Smart Lighting }
\begin{figure}[h]
\centering
\includegraphics[width=1\linewidth,height=0.5\textheight]{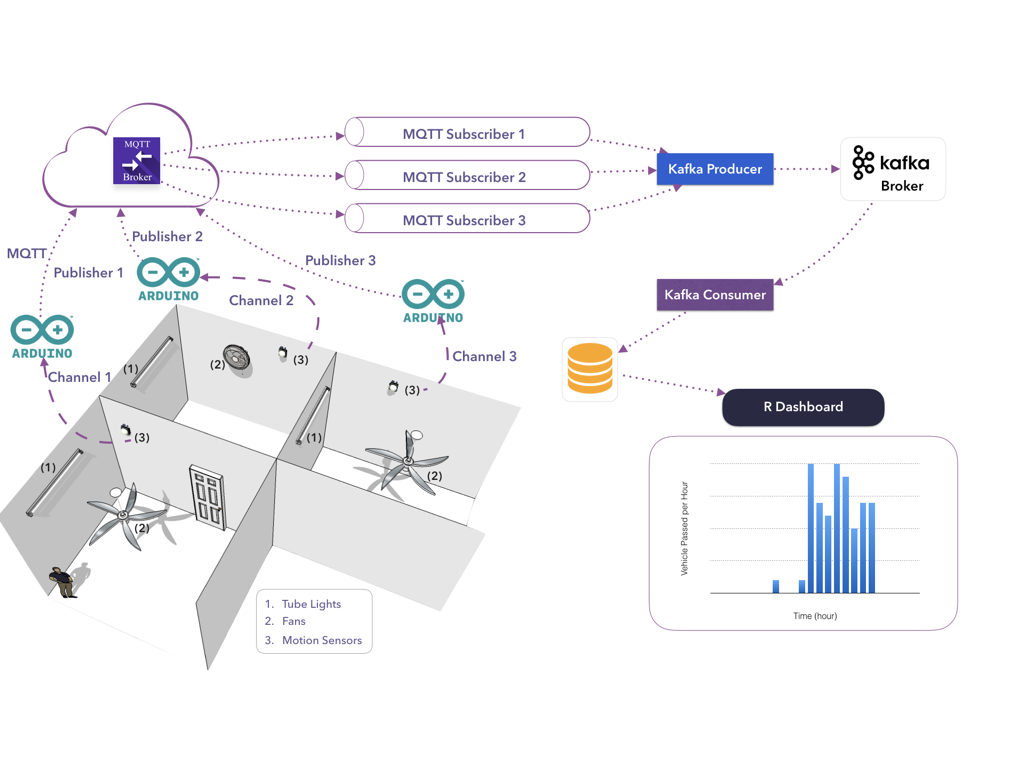}
\caption{Implemented Architecture of Smart Lighting}
\label{fig:smartlighting1}
\end{figure}
\hspace*{7mm} In the second use case of $'$smart lighting$'$,  we have actually implemented a system using PIR motion sensor which senses the motion of object (Especially, warm bodies like animals, people) across its field of view. The functionality of sensor is used with the help of arduino microcontroller. Here, whenever human entry is detected it automatically turns on the light. The light remains on for the next 3 minutes and then automatically it gets turn off without any manual human intervention. These data entries are collected and gets published through MQTT publisher located on arduino module. This published data is sent to MQTT broker at remote location. MQTT subscriber on another remotely located server subscribes to data located in the queue of MQTT broker. These real time collected sensor values are then sent from kafka producer to kafka consumer. Kafka consumer inject this data into the database which further is used for visualization purposes. R platform uses this database to request queries for visualization on dashboard. A package named Shiny is used along with R for interactive dashboard.

\section{Evaluation}\label{sec:evalution}
\subsection{Use Case - 1 : Traffic Scenario}
\hspace*{7mm} With the constant economic development, the number of vehicles are increasing quickly. Road traffic is . Congested driving conditions are happened every now and again on occupied streets in city, which influences people$'$s work, and causes immense misfortune of individuals and society.
In India, road traffic conditions are becoming worse everyday.At the rate of 10.16 percent annually, average number of vehicles in India is growing since last five years ~\cite[NHAI]{ten}. In metropolitan cities like Mumbai flow of vehicles is about 590 vehicles per Km of road stretch ~\cite[Infrastructure, wiki]{eleven}. In this way, it is an earnest need to improve traffic management, including real-time observing vehicles running on the road, making dynamic insights of constant real time traffic flow, acknowledging optimized traffic scheduling.
\begin{figure}[h]
\centering
\includegraphics[width=1\linewidth]{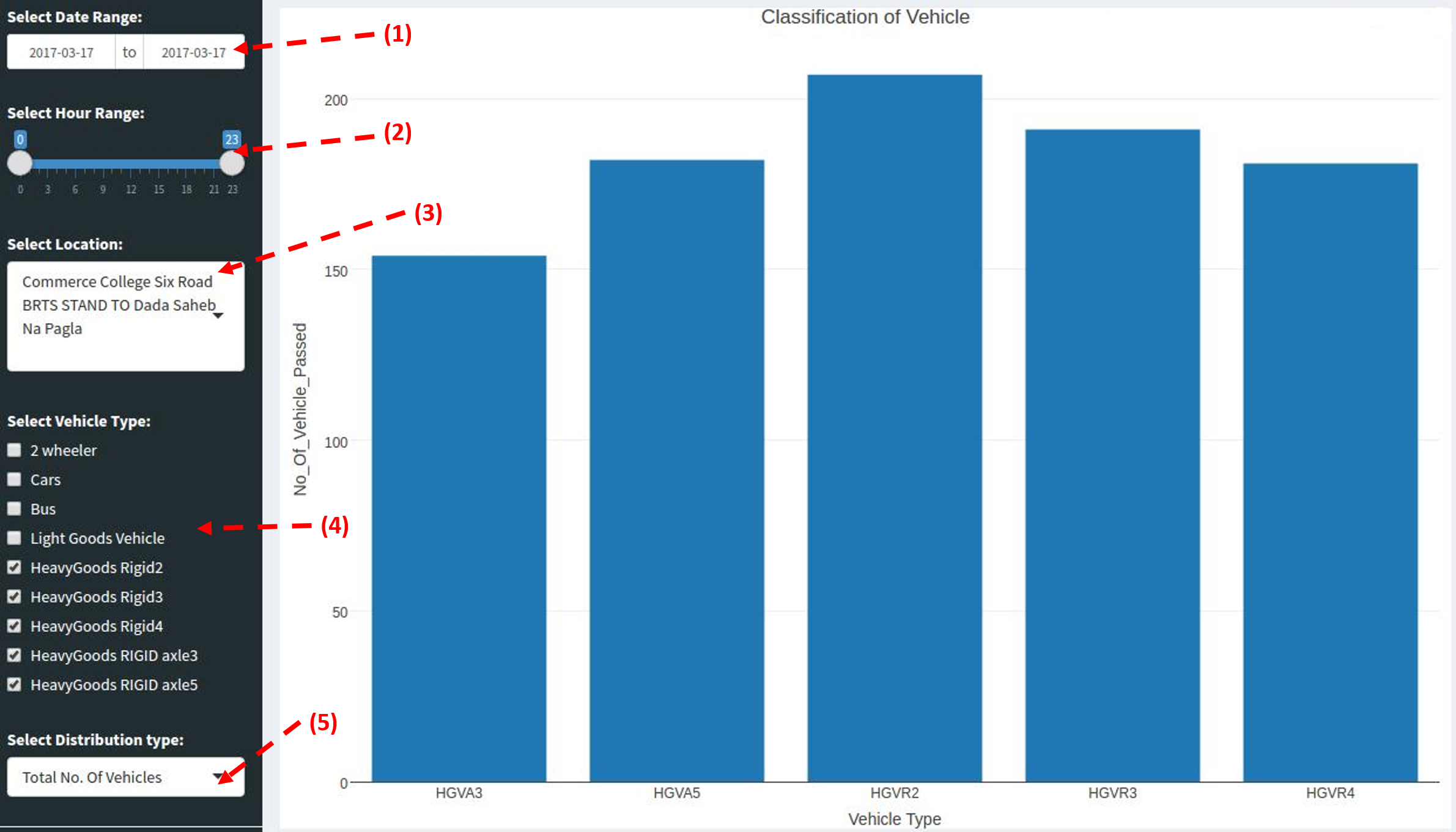}
\caption{Classification of Vehicles by Date and Time. Interactive dashboard with input (1)Date Range (2) Hour range (3) Location (4) Type of Vehicles (5) Distribution Type }
\label{fig:traffic1}
\end{figure}
\\
\hspace*{3mm}Figure ~\ref{fig:traffic1} shows the number of vehicles passed by with their category such as Car, Bus, Two motor vehicles on particular route.  An end user can select parameters mentioned in above figure and filters out the information by specific date, hour-range, location (A particular route), type of vehicles and distribution type such as Total number of Vehicles or Average vehicles per minute. If end user select the current date then the graph will update in real time at every minute. 
\begin{figure}[h]
\centering
\includegraphics[width=1\linewidth]{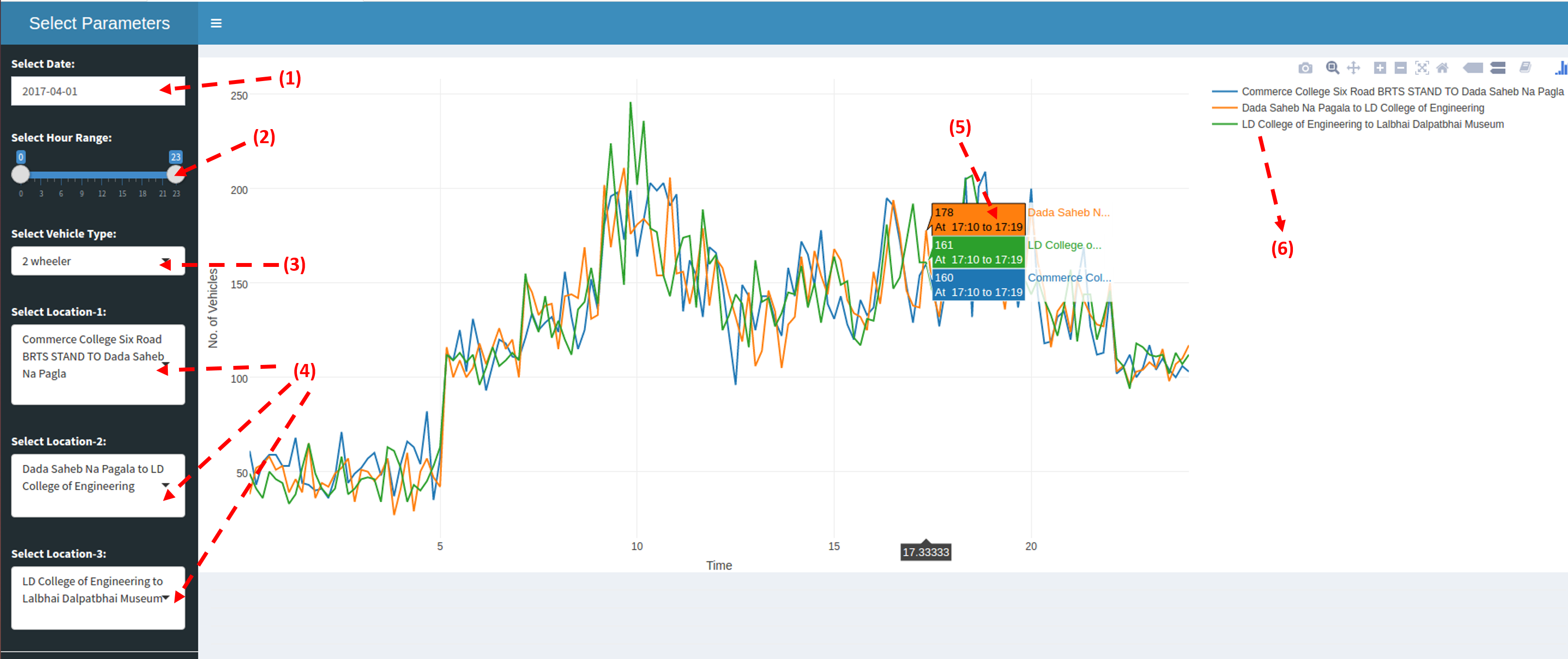}
\caption{Comparison of Traffic Density of different Locations by Date and Time. Interactive dashboard with input elements (1)Date (2) Hour range (3) Type of Vehicles (4) Locations. Plot elements (5)pointer roll over marker and (6) Legend }
\label{fig:traffic2}
\end{figure}
\\\hspace*{7mm} Figure ~\ref{fig:traffic2} shows the comparison of vehicle flow of different routes. An end user can select parameters mentioned in above figure and can filter out the information by date,hour-range,locations and type of vehicles. This real time streaming data is visualized to realize the the density of vehicular traffic on different routes on a given date and time.From this end user can analyze that on which route traffic is high. 
\newpage
\begin{figure}[h]
\centering
\includegraphics[width=1\linewidth]{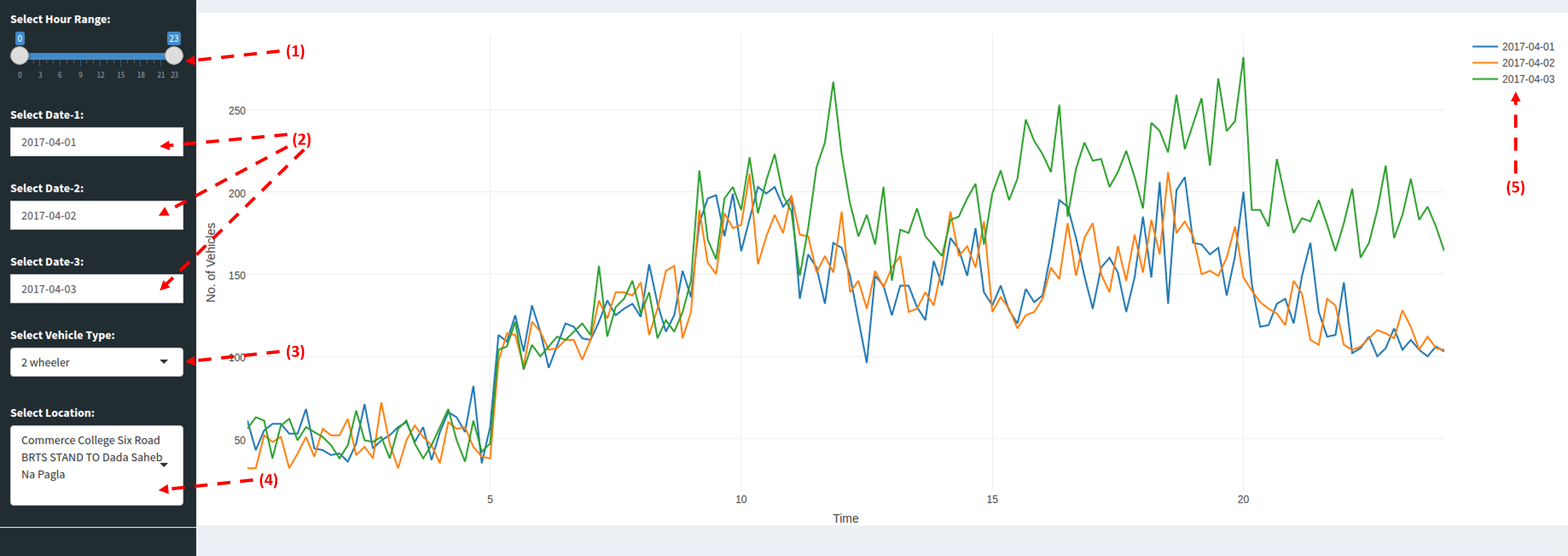}
\caption{Comparison of Traffic density of particular location on different date. Interactive dashboard with input elements (1)Hour range (2)Date (3) Type of Vehicles (4) Locations. Plot elements (5) Legend}
\label{fig:traffic3}
\end{figure}
\hspace*{3mm} Figure ~\ref{fig:traffic3} shows the vehicular density of same route on different dates.Here, end user can see the vehicular traffic condition on different selected dates and for a specific hour range on the same route. An end user can select parameters mentioned in above figure and can filters out the information by date, hour-range and type of vehicles. If end user select the current date then graph will update in real time at every minute.\\\\
\hspace*{7mm} The above generated simulation results can be helpful for civil authorities to make important decisions on designing transport infrastructure, policies and regulations. In addition to this, it can also be useful in future integration of technologies. This knowledge can be taken into  account for setting up the signal timings at cross roads. By scaling this project, more exhaustive data can be collected for detailed visualization which can pertain much advance recommendation and suggestion.\\
\newpage
\subsection{Use Case - 2 : Smart Lighting}
\hspace*{7mm} With the development of economy and urbanization, the rapid growth of infrastructure becomes the prime contributor in energy consumption. The UN assesses that cities are in charge of near 75 percent of worldwide essential energy and 70 percent of worldwide carbon outflows. The circumstance in India is especially serious - peak electricity request in our cities is rising every year but more than 400 million individuals in India are as yet sitting tight for access to dependable source of energy. British Petroleum's Energy Outlook 2035 states that energy demand in India is anticipated that would increment by 132 percent by 2035 while the development in production will be close to 112 percent ~\cite[Hayagrish,2017]{nine}. Efficient utilization of energy is one of the main concerns in the development of smart cities. In order to have a sustainable development, we need to put significant efforts in reducing the consumption of energy. These proposition can be very well promoted in educational institutes. Smart campus is an appealing application in the paradigm of the IoT. The efforts emphasizes on making the facilities more efficient with optimum usage of energy. These efforts are can be recognized as constructing a \textquotedblleft Green campus \textquotedblright.\\
\begin{figure}[h!]
\centering
\includegraphics[width=1\linewidth,height=0.25\textheight]{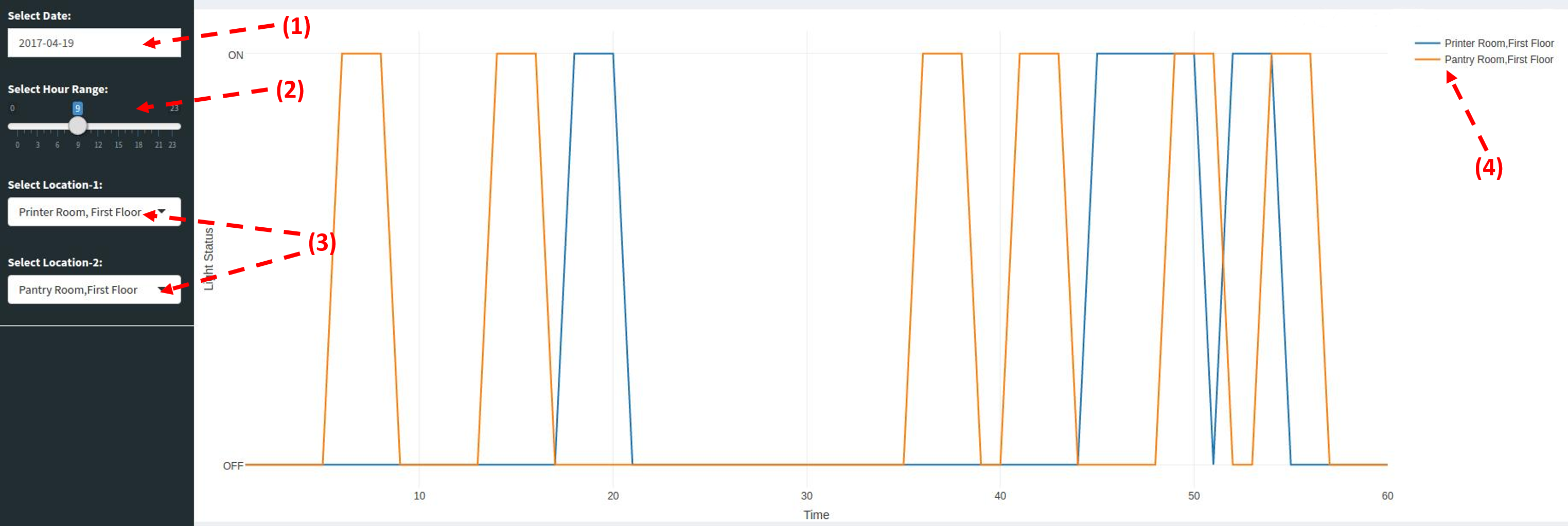}
\caption{Comparision of Enegry consumption of different locations. Interactive dashboard with input elements (1)Date (2)Hour range (3) Locations. Plot elements (4) Legend}
\label{fig:Energybylocation}
\end{figure}
\\\\
\hspace*{3mm} Figure ~\ref{fig:Energybylocation} shows the comparison of energy consumption of two different locations. From this end user can analyze that, in which location energy consumption is more. An end user can select parameters described in above figure and filter out the energy consumption by specific date, time(hour range) and location. If end user select the current date then the graph will update every thirty seconds. 
\begin{figure}[h]
\centering
\includegraphics[width=1\linewidth, height=0.25\textheight]{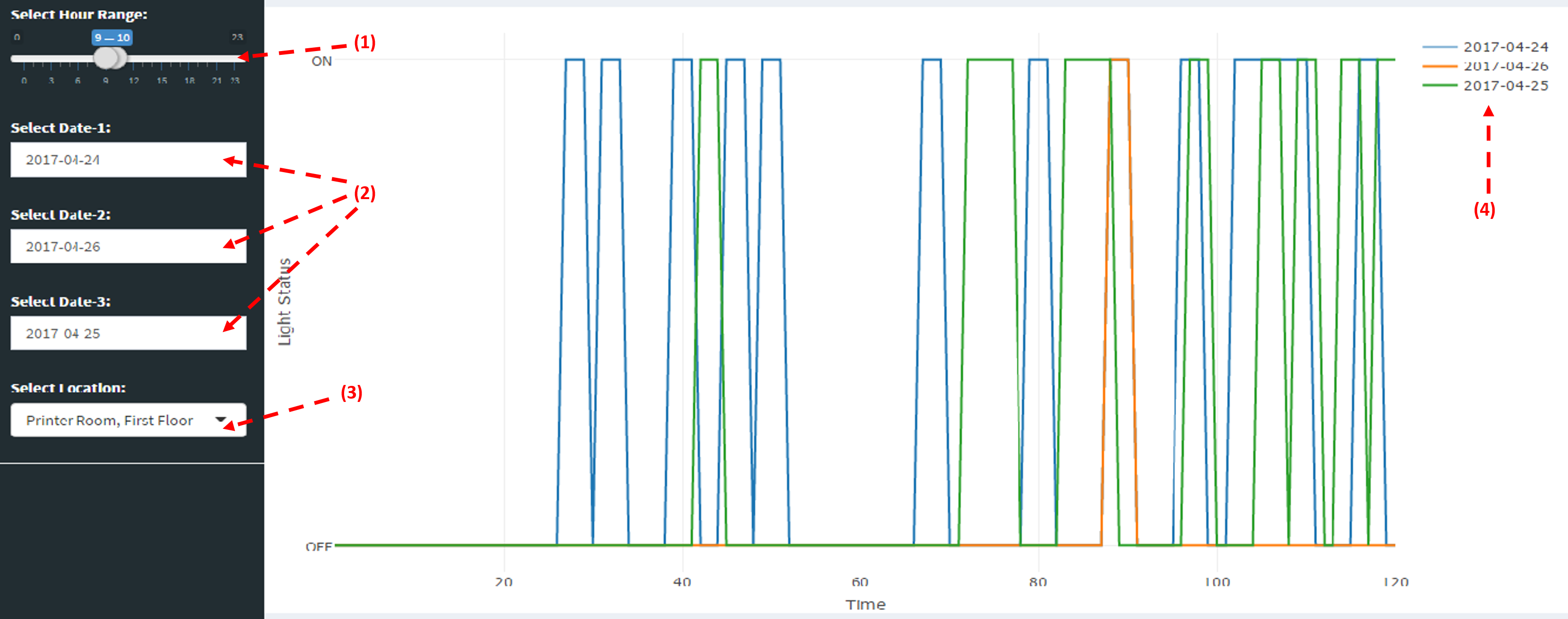}
\caption{Comparison of Energy Consumption of Location on different date. Interactive dashboard with input elements (1)Hour range (2)Date  (3) Locations. Plot elements (4) Legend}
\label{fig:Energybylocation}
\label{fig:123}
\end{figure}
\\
\hspace*{7mm}Figure ~\ref{fig:123} shows the comparison of the energy consumption of one location on different dates. From this you can analyze that, on which date or day energy consumption is more. An end user can select parameters described in above figure and can filter out the energy consumption by specific date, time(hour range) and location. If end user select the current date then the graph will update every thirty seconds. 
\begin{figure}[h]
\centering
\includegraphics[width=1\linewidth]{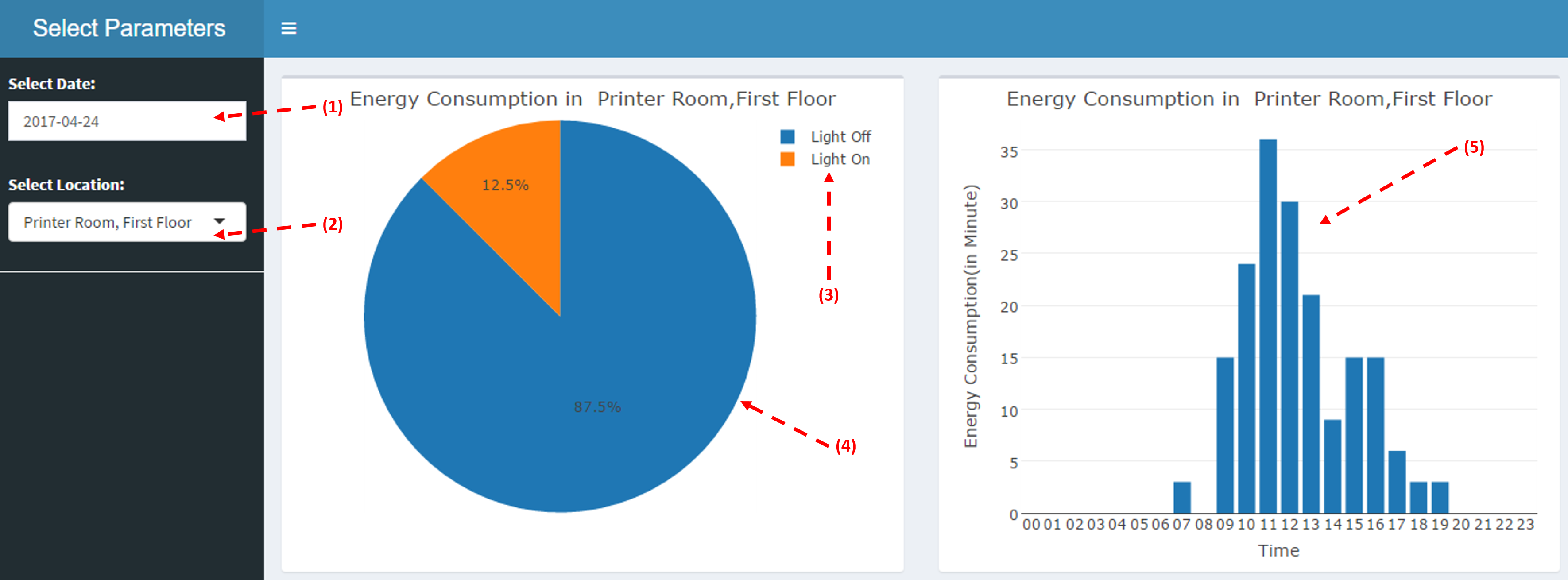}
\caption{Total Energy Consumption of Location. Interactive dashboard with input elements (1)Date (2)Locations. Plot elements (3) Legend for pi chart (4) Pi chart (5) Bar plot}
\label{fig:pibarsmartlight}
\end{figure}
\\\hspace*{7mm} Figure ~\ref{fig:pibarsmartlight}, shows the utilization of energy of a given date in the experimental location. Pie chart shows percentage of how much time energy is consumed on a given date and how much time energy is not consumed. Bar chart shows the energy consumption of a given date hour wise. From that end user can analyze that, in which hours more numbers of people coming to that location. If end user select the current date then the graph will update every thirty seconds. \\\\
\hspace*{7mm} The above simulated results shows the energy consumption visualization and how smart lighting is reducing the energy consumption significantly. The more scaled up version of this project can include such motion sensors in all the rooms restricting the over utilization of energy and leading to smart building in terms of energy consumption. In addition to this, data analytics on the energy consumption dataset can cater well to identify the patterns in  the usage of energy and thus, can lead to smart and efficient management of energy.\\

\section{Related work}\label{sec:relatedwork}
\hspace*{7mm} One of the contemporary work in the field of IoT testbed is WoTT (Web Things testbed) for the testing and designing real world innovative IoT products and application that solves the problem in different areas such as healthcare, transportation, security, etc. The WoTT is an innovative and heterogeneous Web of things based testbed which leverages developer community to design and assess novel applications and services in real IoT surroundings with ease and to test human-object interactivity that plays crucial role in widening the horizon of IoT use. Similar to Web of Things Testbed, our testbed is very well suited for the above objectives as its architecture is primarily built on standard network interfaces and protocols, without proprietary solutions that would injure the interoperability among different modules.WoTT (Web of things testbed)$'$s is aimed to :hide low-level implementation details, enhance network self-configuration with minimum human interaction in the loop, simultaneously manage different and multiple protocols, frameworks and platforms, and provide a platform for the design and testing of human$-$object interaction patterns ~\cite[Belli, Laura]{first}. \\\\
\hspace*{7mm} The primary challenges experienced in WoTT deployment is related to its design. For instance, the precise definition of different modules and how they functions, how to present various resources and their interaction or relationships through appropriate hypermedia and preserve compatibility among different standards. The attempts aimed to the IoT Hub, WoTT design, and the use of common standards have significantly reduce the challenges involved in the deployment process, which makes the integration of different modules uncomplicated in spite of their heterogeneity. Not only, WoTT simplifies communication among IoT nodes, but also forms a uniform super entity that provides advance functions which provides more facilities than just the union of its components features. In order to leverage this super-entity position, WoTT utilizes varied communication technologies in the IoT hub to incorporate and link together various networks into a single IP network. IoT hub is not only simplifies interaction and interoperability, but it also extends the reach of IoT network and increases its capacity and functionalities by simplifying and underlying complicated and salient tasks as service discovery and routing ~\cite[Belli, Laura]{first}.\\\\ 
\hspace*{7mm} Case-1: IoT-Lab - IoT lab successfully identifies the potential of crowdsourcing as an extension to conventional IoT infrastructure.  It is large scale infrastructure consisting more than 2700 wireless sensors nodes which are spread across six different cities in France. This system is used to test link and network later protocols and to collect performance results (Such as energy consumption and packet delivery ratio). Here, the project proposed a novel approach for testing and implementing solution for real world problems with the participation of crowd in the process.  Here, user participates in the experiment by contributing with sensory data and knowledge. The objective is to engage the crowd in achieving goals of innovative products, process optimization and problem solving in different areas and industries such as education, healthcare, IT and finance, design, entrepreneurship, social enterprise and many more. In this approach, the crowd generally contributes with knowledge, time, feedback, ideas, resources, or even funding to a project ~\cite[Fernandes, Jo.o]{second}.
\\\hspace*{7mm} In this crowd driven approach, the crowd proposes a potential experiment for IoT lab platform. Users work as sensing nodes and provide information through the smartphone application. These ideas and information are collection and the crowd assess this information or ideas. The chosen idea is implemented by the researchers and again evaluated by the crowd. This iterative process continues and generates more impactful research and results.\\\\
\hspace*{7mm} The paper proposed a unique model of the Testbed as a Service (TBaaS). In this model, the integrated testbeds are accessible by cloud technologies in transparent and uniform way. All these testbed will function as single IoT meta-testbed delivering a TBaaS cloud layer with help of underlying mechanisms and APIs which meet the requirements and the architecture design of IoT lab platform.  Similar to TBaaS, in our proposed idea of a testbed the integrated virtualized resources will be available through range of technologies that are bolstered by most of the contemporary systems and devices. In addition to that, it will provide new business models for better development of Testbed experiments ~\cite[Fernandes, Joao]{second}.\\\\ 
Case-2:  SmartSantander- There are 20,000 sensing nodes deployed across Europe for sensing parameters (temperature, CO, noise, light, car presence, etc.) (Environmental data collection).\\\\
\hspace*{7mm} One of the key consideration while building an IoT testbed is its scope. With the concern of technological scope, if a testbed addresses only a specific type of technology then it is considered as a single domain, e.g. wireless sensors network or RFID devices. While, a multi domain incorporates various technologies into a common experimental facility.  A multi-domain is progressively important in dealing with requirements of IoT heterogeneity. In addition to this, testbed can also be realize either as indoor (Motelab) or outdoor (CitySense, Oulu smart city, SmartSantander). Clearly, the variety in domains, influences the choice of software mechanisms and hidden hardware architecture.  In indoor installations, it provides electricity and cabling for testbed control and management which makes it  easier and available.  Contrary to that, outdoor testbeds have to be dependent on wireless connections which requires extra mechanisms for reliability. They also need be safeguarded against spiteful physical access and other threats which don’t generally occur often in controlled setups.  Moreover, testbeds can be realized as permanent setups that are consistently available or portable ones, which are generally deployed for measurement campaigns ~\cite[Gluhak, Alexander]{third}.
\\\hspace*{7mm}Furthermore, testbed can differ in terms of application domain they are targeted to investigate. Generic testbeds allow experimentation research of communication protocols at different layers of protocol stack or application level algorithms that are not dependent on a specific application or service. Recently, the emerging idea of domain specific testbeds allow the analysis of applications and services in real environment from the perspective of an end-user.The proposed testbed is generic, multi-domain, combination of indoor and outdoor control and management end-to-end system.
\\\hspace*{7mm} SmartCampus is a Testbed deployed in real world office with spanning entire building, overcomes existing testbeds by improving important dimensions like realism (Real end user involvement) and device heterogeneity for experimenting lab based IoT testbeds. SmartCampus have 200 IoT nodes installed on each desk of work area sending data of noise, temperature, light, and motion to the main cloud server and they notify the user about energy consumption on smartphone and allow the user to turn off device remotely via an android app. For 200 installed IoT nodes 100 Gateways are installed to remotely program IoT node ~\cite[Nati, Michele]{fourth}.\\\\
\hspace*{7mm} In results, SmarCampus testbed has developed an interactive visualization of an activity heatmap show the no. of active devices at a particular time for a data of a week.SmartCampus includes Social graph and identified user/Device highest centrality during the day~\cite[Nati, Michele]{fourth}.

\section{CONCLUSION AND FUTURE WORK}\label{sec:conclusion}
\hspace*{7mm} A testbed is a useful solution for checking the feasibility of a concept or an idea. Not just this, such testbeds helps the research fraternity in solving the real world problems with the state-of-the-art technologies at a rapid pace. The existence of such open source projects helps in reducing significant efforts in the configuration and integration of multiple technologies and frameworks. Undoubtedly, there are issues pertaining to  the authenticity of datasets for research and analysis purposes. However, the cases for which the authenticity of data is not in question,the proposed system can perform well for the required analysis. The paper described two use cases: 1.Traffic scenario and 2. Smart lighting.Though the scale of implementation was small, we achieved significant success from the whole system point of view. This can be scaled further with more number of sensing nodes deployed on the required locations and leveraging the infrastructure for the purpose of storage and processing. Regarding the extensibility of testbed, it can be made accessible through cloud technologies for further contribution by developer community.

\end{document}